\documentclass{jfm}

\usepackage{amsmath}
\usepackage{amsfonts}
\usepackage{bm}
\usepackage[mathcal]{euscript}
\usepackage{graphicx}
\usepackage{psfrag}
\usepackage{subfigure}

\ifCUPmtlplainloaded \else
  \checkfont{eurm10}
  \iffontfound
    \IfFileExists{upmath.sty}
      {\typeout{^^JFound AMS Euler Roman fonts on the system,
                   using the 'upmath' package.^^J}%
       \usepackage{upmath}}
      {\typeout{^^JFound AMS Euler Roman fonts on the system, but you
                   dont seem to have the}%
       \typeout{'upmath' package installed. JFM.cls can take advantage
                 of these fonts,^^Jif you use 'upmath' package.^^J}%
      }
  \else
  \fi
\fi

\ifCUPmtlplainloaded \else
  \checkfont{msam10}
  \iffontfound
    \IfFileExists{amssymb.sty}
      {\typeout{^^JFound AMS Symbol fonts on the system, using the
                'amssymb' package.^^J}%
       \usepackage{amssymb}%
       \let\le=\leqslant  
       \let\ge=\geqslant  
      }{}
  \fi
\fi

\ifCUPmtlplainloaded \else
  \IfFileExists{amsbsy.sty}
    {\typeout{^^JFound the 'amsbsy' package on the system, using it.^^J}%
     \usepackage{amsbsy}}
    {}
\fi

\newcommand{\goodgap}{%
	\hspace{\subfigtopskip}%
	\hspace{\subfigbottomskip}}

\newcommand{\ie}{\textit{i.e.}}

\newcommand{\mathnotation}[2]{\newcommand{#1}{\ensuremath{#2}}}

\newcommand{\Lnorm}[2]{\l\lVert#1\r\rVert_{#2}}

\renewcommand{\l}{\left}			
\renewcommand{\r}{\right}			
\mathnotation{\pd}{\partial}			
\mathnotation{\ldef}{\mathrel{\raisebox{.069ex}{:}\!\!=}}
\mathnotation{\rdef}{\mathrel{=\!\!\raisebox{.069ex}{:}}}
\mathnotation{\dint}{\,{\mathrm{d}}}		

\mathnotation{\grad}{\nabla}			
\mathnotation{\lapl}{\Delta}			

\mathnotation{\Torus}{{\mathbb{T}}}		
\mathnotation{\Unit}{{\mathbb{I}}}		
\mathnotation{\sdim}{d}				

\renewcommand{\time}{t}				
\mathnotation{\Time}{T}
\mathnotation{\pdt}{\partial_\time}		
\mathnotation{\xc}{x}				
\mathnotation{\yc}{y}				
\mathnotation{\xv}{{\bm{\xc}}}			
\mathnotation{\yv}{{\bm{\yc}}}			
\mathnotation{\uc}{u}				
\mathnotation{\uv}{\bm{\uc}}			
\mathnotation{\Upsv}{{\bm{\Upsilon}}}		

\mathnotation{\Diff}{\kappa}			
\mathnotation{\Diffm}{\eta}			
\mathnotation{\vardiss}{\epsilon}		
\mathnotation{\ssc}{s}				
\mathnotation{\Ssc}{S}				
\mathnotation{\lavg}{\langle}
\mathnotation{\ravg}{\rangle}
\mathnotation{\Pe}{\mathit{Pe}}			
\mathnotation{\Sr}{\mathit{Sr}}			
\mathnotation{\cci}{c_1}
\mathnotation{\ccii}{c_2}
\mathnotation{\ccbi}{\bar c_1}
\mathnotation{\ccbii}{\bar c_2}
\mathnotation{\DiffEq}{\Diff_{\mathrm{eq}}}	
\mathnotation{\DiffEff}{\Diff_{\mathrm{eff}}}	

\newcommand{\NS}{Navier--Stokes}

\newcommand\etal{\mbox{\textit{et al.}}}

\newcommand\eg{e.g.\ }

\begin{document}

\title[Bound on Mixing Efficiency]{A Bound on Mixing Efficiency for the
  Advection--Diffusion Equation}
\author[Jean-Luc Thiffeault, Charles R. Doering and John D. Gibbon]%
{J\ls E\ls A\ls N\ls -\ls L\ls U\ls C\ns%
T\ls H\ls I\ls F\ls F\ls E\ls A\ls U\ls L\ls T,$^1$\break
C\ls H\ls A\ls R\ls L\ls E\ls S\ns  R.\ns D\ls O\ls E\ls R\ls I\ls N\ls G%
$^2$
\and J\ls O\ls H\ls N\ns D.\ns G\ls I\ls B\ls B\ls O\ls N$^1$}
\affiliation{$^1$Department of Mathematics, Imperial College London, SW7 2AZ,
UK\\[\affilskip]
$^2$Department of Mathematics and Michigan Center for Theoretical
Physics, University of Michigan, Ann Arbor, MI 48109-1109, USA}

\pubyear{2004}
\volume{521}
\pagerange{105--114}
\date{?? and in revised form ??}
\setcounter{page}{1}

\maketitle

\begin{abstract}
An upper bound on the mixing efficiency is derived for a passive scalar under
the influence of advection and diffusion with a body source.  For a given
stirring velocity field, the mixing efficiency is measured in terms of an
\emph{equivalent diffusivity}, which is the molecular diffusivity that would
be required to achieve the same level of fluctuations in the scalar
concentration in the absence of stirring, for the same source distribution.
The bound on the equivalent diffusivity depends only on the functional
``shape'' of both the source and the advecting field.  Direct numerical
simulations performed for a simple advecting flow to test the bounds are
reported.
\end{abstract}

\section{Introduction}
\label{sec:intro}

In this work we apply some recent developments in the analysis of the \NS\
equations (\cite[Doering \& Foias 2002]{Doering2002}) to mixing and the
advection--diffusion equation.  Mixing phenomena are ubiquitous with
applications in atmospheric science, oceanography, chemical engineering, and
microfluidics, to name a few.  Here we focus on the generic problem of the
advection--diffusion equation with a source that replenishes the variance of
the passive scalar.  The stirring is effected by a specified velocity field,
which may or may not be turbulent.  Our analysis of an idealised model lends
mathematical precision and rigour to conventional scaling arguments often
invoked for these kinds of problems.

For the passive scalar, complicated behaviour---and efficient mixing---is
often observed even for laminar velocity fields.  This is the well-known
effect of chaotic advection (\cite[Aref 1984]{Aref1984}; \cite[Ottino
1989]{Ottino}).  Thus we can choose the \emph{stirring} (the advecting
velocity field) to be any divergence-free, possibly time-dependent flow field.
The mixing efficiency then depends on specific properties of the stirring
field as well as the manner in which the scalar concentration is injected,
which is exactly what would be expected.  The bound on mixing efficiency
derived in this paper has that feature: it depends on the stirring field and
the source distribution.  This is very helpful as it allows for comparison of
the relative effectiveness of various stirring scenarios for, say, a specified
source.  The bounds we obtain are also valid for turbulent flows, as we make
no assumptions the smoothness of the stirring field.  A recent study by
\cite{Schumacher2003} has produced bounds on the derivative moments of the
concentration field; here we shall focus on the undifferentiated quadratic
moment.  As will become evident, the methods of this paper can also be
extended to produce bounds on derivatives of the concentration field.


\section{System Description}

The advection--diffusion equation for the concentration~$\theta(\xv,\time)$ of
a passive scalar is
\begin{equation}
  \pdt\theta + \uv\cdot\grad\theta = \Diff\,\lapl\theta + \ssc\,,
  \label{eq:AD}
\end{equation}
where~$\Diff$ is the molecular diffusivity and~$\ssc(\xv,\time)$ is a source
function with zero spatial mean.  The domain we consider is a periodic box of
side~$L$, i.e., ~$\xv\in\Torus^\sdim$, the~$\sdim$-dimensional torus.  The
velocity field $\uv(\xv,\time)$ could be obtained by solving \NS\ or some
other set of equations, but here we shall simply consider it to be an
arbitrary~$L^2$ divergence-free vector field.  Hence without loss of
generality we may take the solution $\theta(\xv,\time)$ to be spatially mean
zero at all times.

Variations in the source term in~\eqref{eq:AD} maintain the inhomogeneity of
the concentration field.  The stirring term may lead to the formation of sharp
gradients of concentration that then enhance the effect of molecular
diffusion.  For definiteness we assume that both the source and the stirring
act on a comparable scale,~\hbox{$\ell\le L$}.  Because of
periodicity,~$L/\ell$ is an integer.  We introduce these two distinct scales
in order to be able to consider the infinite volume limit,~$L \rightarrow
\infty$ at fixed $\ell$, for the final results.

We shall use the \emph{fluctuations} in the concentration as a useful measure
of the degree of well-mixedness, as has long been the practise (\eg
\cite[Danckwerts 1952]{Danckwerts1952}; \cite[Edwards, Sherman \& Breidenthal
1985]{Edwards1985}; \cite[Rehab, Antonia, Djenidi \& Mi 2000]{Rehab2000}).  To
characterise the fluctuations in~$\theta$, we use the variance,
\begin{equation}
  \Theta^2 \ldef \l\lavg 
  L^{-\sdim}\Lnorm{\theta}{L^2(\Torus^\sdim)}^2\r\ravg,
  \label{eq:vardef}
\end{equation}
of the spatially mean-zero concentration.  The angle
brackets~$\lavg\cdot\ravg$ denote a long-time average, which we will assume
exists for the quantities of interest, and~$\Lnorm{\cdot}{L^2(\Torus^\sdim)}$
is the~$L^2$ norm on~$\Torus^\sdim$.  As control parameters we use the
variance of the source and a measure of the kinetic energy density of the
stirring field,
\begin{equation}
  \Ssc^2 \ldef \l\lavg L^{-\sdim}\Lnorm{\ssc}{L^2(\Torus^\sdim)}^2\r\ravg,
  \quad
  U^2 \ldef \l\lavg L^{-\sdim}\Lnorm{\uv}{L^2(\Torus^\sdim)}^2\r\ravg.
  \label{eq:SUdef}
\end{equation}
Thus,~$\Theta$,~$\Ssc$, and~$U$ are spatio-temporal averages respectively of
fluctuations in the scalar concentration~$\theta(\xv,\time)$, the
source~$\ssc(\xv,\time)$, and the fluid velocity~$\uv(\xv,\time)$.  An
efficient mixing configuration would have small~$\Theta$ for a given $\Ssc$
and~$U$, indicating a steady-state with small variations in the concentration.
In general we expect that increasing~$U$ at fixed~$\Ssc$ should
decrease~$\Theta$, for this represents more vigorous stirring, while
increasing~$\Ssc$ at fixed~$U$ should augment~$\Theta$.  We will show in this
paper that~$\Theta$ has a lower bound proportional to~$\Ssc\ell/U$, so that a
source with large fluctuations necessarily produces a poorly mixed state
unless~$U$ is increased sufficiently.

In order to keep track of the effects of the amplitudes of the source
variation and stirring intensity and their characteristic length scales
independently from the influence of the particular ``shapes'' of the input and
mixing functions, we decompose~$s$ and~$\uv$ into the dimensional amplitudes
($\Ssc$ and $U$) and dimensionless shape functions ($\Phi$ and $\Upsv$)
according to
\begin{equation}
  \ssc(\xv,\time) = \Ssc\,\Phi(\xv/\ell,\time/\tau),\quad
  \l\lavg L^{-\sdim}\Lnorm{\Phi}{L^2(\Torus^\sdim)}^2\r\ravg = 1,
  \label{eq:fscdecomp}
\end{equation}
and
\begin{equation}
  \uv(\xv,\time) = U\,\Upsv(\xv/\ell,\time/\tau),\quad
  \l\lavg L^{-\sdim}\Lnorm{\Upsv}{L^2(\Torus^\sdim)}^2\r\ravg = 1,
  \label{eq:uvdecomp}
\end{equation}
where~$\tau$ is an appropriate time scale characterising the source and
stirring.  Of course either or both may be time-independent, but in any case
we presume periodicity or statistical stationarity with identifiable periods
or relaxation times.

\section{The Bounds}
\label{sec:bounds}

Now consider an arbitrary smooth (dimensionless) spatially periodic
function~$\Psi(\xv/\ell,\time/\tau)$ normalised such that
\begin{equation}
  \l\lavg{L^{-\sdim}}\int_{\Torus^\sdim}
  \Psi(\xv/\ell,\time/\tau)\,\Phi(\xv/\ell,\time/\tau)
  \dint^\sdim\xc\r\ravg = 1,
  \label{eq:Psinorm}
\end{equation}
For example because of the normalisation in~\eqref{eq:fscdecomp}, $\Psi=\Phi$
could be a possible choice if it is sufficiently smooth.
Multiply~\eqref{eq:AD} by~$\Psi$ and space-time average.
Using~\eqref{eq:fscdecomp} and~\eqref{eq:Psinorm} and integrating by parts, we
may express~$\Ssc$ as
\begin{equation}
  \Ssc = -\l\lavg {L^{-\sdim}}\int_{\Torus^\sdim}
  \l(\pdt\Psi + \uv\cdot\grad\Psi
  + \Diff\,\lapl\Psi\r)\,\theta\dint^\sdim\xc\r\ravg.
  \label{eq:Sscsolved}
\end{equation}
Note that the operator acting on~$\Psi$ in~\eqref{eq:Sscsolved} is the adjoint
of the advection--diffusion operator, which suggests how the method can be
generalised to other linear operators with a body source (e.g., the magnetic
induction operator of dynamo theory (\cite[Childress \& Gilbert 1995]{STF})).

The Cauchy--Schwartz inequality implies the bound
\begin{equation}
  \Ssc \le \l\lavg L^{-\sdim}\Lnorm{
  \pdt\Psi + \uv\cdot\grad\Psi
  + \Diff\,\lapl\Psi}{L^2(\Torus^\sdim)}^2\r\ravg^{1/2}
  \Theta.
\end{equation}
Then substituting the scaled variables~$\Time=\time/\tau$ and~$\yv=\xv/\ell$
and using~\eqref{eq:uvdecomp}, we have
\begin{equation}
  \Ssc \le \frac{U\Theta}{\ell}\l\lavg \Lnorm{\Omega}{L^2(\Unit^\sdim)}^2
  \r\ravg^{1/2}
  \label{eq:Sscbound}
\end{equation}
where~$\Unit=[0,1]$ is the unit torus and
\begin{equation}
  \Omega(\yv,\Time) \ldef - \Sr\,\pd_{\Time}\Psi(\yv,\Time) -
  \Upsv(\yv,\Time)\cdot\grad_\yv\Psi(\yv,\Time)
  + \frac{1}{\Pe}\,(-\lapl_\yv\Psi(\yv,\Time)).
\end{equation}
Here the P\'{e}clet number is~$\Pe=U\ell/\Diff$.  If the velocity field is
time-dependent with timescale~$\tau$, the dimensionless number~\hbox{$\Sr
\ldef \ell/U\tau$} may be regarded as a Strouhal number; in any case, we shall
refer to it as the Strouhal number even if the timescale~$\tau$ is unrelated
to~$\uv$.

In principle inequality~\eqref{eq:Sscbound} could be sharpened by
varying~$\Psi$ to provide as tight a bound as possible, as performed
by~\cite{Doering2003} for the power consumption rate in the \NS\ equations.
We will not pursue that direction here; rather we will produce explicit limits
via simple estimates.

Applying the Minkowski inequality to~\eqref{eq:Sscbound}, we see that
\begin{equation}
  \Ssc \le \frac{U\Theta}{\ell}\l(\cci + \Pe^{-1}\,\ccii\r)
  \label{eq:Sscbound2}
\end{equation}
where
\begin{subequations}
\begin{align}
  \cci &\ldef
  \l\lavg\Lnorm{
    \Sr\,\pd_{\Time}\Psi
    + \Upsv\cdot\grad_\yv\Psi}{L^2(\Unit^\sdim)}^2\r\ravg^{1/2},
  \label{eq:c1def}\\
  \ccii &\ldef \l\lavg\Lnorm{\lapl_\yv\Psi}{L^2(\Unit^\sdim)}^2\r\ravg^{1/2}.
  \label{eq:c2def}
\end{align}
\label{eq:c12def}%
\end{subequations}
are dimensionless constants, independent of~$\Pe$ and $\Theta$.  The
constant~$\cci$ depends on dimensional quantities only through the Strouhal
number; it also depends explicitly on the stirring shape-function~$\Upsv$.
Note also that the function~$\Psi$ depends indirectly on the source
shape-function~$\Phi$ through its normalisation~\eqref{eq:Psinorm}, so that
both the source and stirring shapes enter the bound.  The constant~$\ccii$
controls the diffusive part while $\Diff$ only enters through the P\'eclet
number in~\eqref{eq:Sscbound2}.  We still have the freedom to choose~$\Psi$ to
optimise~$\cci$ for a particular problem, that is, for particular source and
stirring shapes~$\Phi$ and~$\Upsv$.
 
For small~$\Pe$, we can focus on the~$\ccii$ term in~\eqref{eq:Sscbound2} and
obtain the bound~\hbox{$\Ssc \le \ccii\Theta\Diff/\ell^2$}.  As we increase
the source amplitude~$\Ssc$, holding the other parameters constant, the
time-averaged variance~$\Theta^2$ must eventually increase.  An increase in
the variance implies that the scalar is more poorly mixed.  There is no
avoiding this unless we increase~$\Diff$ or decrease the scale of the
source~$\ell$: the efficiency of mixing is intrinsically related to the
diffusive mixing rate on the scale of the source variance injection,
\ie,~$\Diff/\ell^2$.

For large~$\Pe$, the more interesting limit for many physical problems, we
focus on the~$\cci$ term in~\eqref{eq:Sscbound2} to get the bound \hbox{$\Ssc
\le \cci U\Theta/\ell$}.  (This is true for sufficiently smooth~$\Psi$.)  As
we increase the source amplitude~$\Ssc$, holding everything else constant, the
bound~\eqref{eq:Sscbound2} again implies we must eventually see an increase in
the steady-state variance,~$\Theta^2$.  However, unlike the small~$\Pe$ case,
we can now (potentially) postpone that increase by raising~$U$, \ie, by
stirring more vigorously.  The exact value of~$\cci$ depends on both
shape-functions, but~\eqref{eq:c1def}, where~$\cci$ is defined, can be broken
up by the Minkowski and H\"older inequalities to give
\begin{equation}
  \cci \le \Sr\l\lavg\Lnorm{\pd_{\Time}\Psi}{L^2(\Unit^\sdim)}^2\r\ravg^{1/2}
    + \sup_{\yv,\time}\l\lvert{\grad_\yv\Psi}\r\rvert,
    \label{eq:absbound}
\end{equation}
which is uniform in the shape of the stirring.  The large~$\Pe$ bound has the
nice feature of being independent of the diffusivity~$\Diff$, a result
expected to hold for the passive scalar under turbulent or chaotic mixing.
However, the linear scaling with~$U$ in Eq.~\eqref{eq:Sscbound2} is not always
appropriate, as will be seen in Section~\ref{sec:sineflow} for the specific
case we have studied numerically.  Note also that the
bound~\eqref{eq:absbound} on~$\cci$ still involves the velocity for
time-dependent~$\Psi$ through the Strouhal number.  If it is possible to
choose~$\Psi$ to be time-independent and still satisfy the normalisation
condition~\eqref{eq:Psinorm}---for example if the source~$\ssc$ is
time-independent---then we have the bound
\begin{equation}
  \cci \le \sup_\yv\l\lvert{\grad_\yv\Psi}\r\rvert,
    \label{eq:absbound2}
\end{equation}
which is satisfied for all possible stirring flows (\ie, any shape
function~$\Upsilon$) independently of~$U$.

We can also derive a lower bound for~$\Ssc$.  The average variance dissipation
rate,~$\vardiss$, satisfies
\begin{equation}
  \vardiss = \l\lavg\Diff\,L^{-\sdim}\Lnorm{\grad\theta}{L^2(\Torus^\sdim)}^2
  \r\ravg
  = \l\lavg{L^{-\sdim}}\int_{\Torus^\sdim}
  \ssc(\cdot,\xv)\,\theta(\cdot,\xv)\dint^\sdim\xc\r\ravg
  \label{eq:vardissdef}
\end{equation}  
where we have used the fact that $\Lnorm{\theta}{L^2(\Torus^\sdim)}$ is
uniformly bounded in time, which is true under the physical assumption that
$\Lnorm{s}{L^2(\Torus^\sdim)}$ is itself uniformly bounded in time.  By using
Poincar\'{e}'s inequality in~\eqref{eq:vardissdef} we
have~\hbox{$(\vardiss/\Diff)^{1/2} \ge\ (2\pi\Theta)/L$}, and the
Cauchy--Schwartz inequality along with the normalisation of~$\Phi$
in~\eqref{eq:fscdecomp} gives~\hbox{$\vardiss \le \Ssc\,\Theta$}.  Together
these give the bound
\begin{equation}
  \Ssc \ge (2\pi/L)^2\,\Diff\,\Theta.
  \label{eq:Ssclowerbound}
\end{equation}
This lower bound reflects that no matter how we stir---or if we do not
stir---there is still some diffusive dissipation of the scalar variance.  The
lower bound~\eqref{eq:Ssclowerbound} also implies that if there is any
variance~$\Theta^2$ present at the steady state, then it must be due to some
minimum amount of amplitude of the source; stirring alone can never generate
scalar variance in this kind of model.

The consequence of the two bounds for~$\Ssc$ is that larger~$\Theta$ must
eventually imply large~$\Ssc$ (from~\eqref{eq:Ssclowerbound}, at fixed~$\Diff$
and~$L$), but large~$\Ssc$ does not necessarily imply large~$\Theta$, as the
difference can be made up by a large~$U$ in~\eqref{eq:Sscbound2}.  This is
what makes enhanced mixing possible.

We may also estimate the typical size of small scales in the scalar field.
Using the bound~\hbox{$\Ssc \ge \vardiss/\Theta$} mentioned above, we can
transform \eqref{eq:Sscbound} and~\eqref{eq:Sscbound2} into upper bounds
for~$\vardiss$, viz.
\begin{equation}
  \l(\frac{2\pi}{L}\r)^2\Diff\,\Theta^2 \le
  \vardiss \le
  \frac{U\Theta^2}{\ell}\l(\cci + \Pe^{-1}\,\ccii\r),
\end{equation}
where the lower bound is obtained via Poincar\'{e}'s inequality.  If we define
a scalar dissipation scale~$\lambda$,
\begin{equation}
  \lambda^{-2} \ldef
  \frac{\lavg\Lnorm{\grad\theta}{L^2(\Torus^\sdim)}^2\ravg}
       {\lavg\Lnorm{\theta}{L^2(\Torus^\sdim)}^2\ravg}
  = \frac{\vardiss}{\Diff\,\Theta^2}\,,
\end{equation}
(the Batchelor scale [\cite[Batchelor 1959]{Batchelor1959}], an analog of the
Taylor microscale for the \NS\ equations) then
\begin{equation}
  L/(2\pi) \ge \lambda \ge \ell\l(\cci\,\Pe + \ccii\r)^{-1/2}.
\end{equation}
For large~$\Pe$, the smallest possible size of this dissipation scale is
proportional to~$\Pe^{-1/2}$, a standard theoretical estimate (\cite[Childress
\& Gilbert 1995]{STF}).

\section{Mixing Efficiency and Equivalent Diffusivity}

As a physically meaningful measure of mixing efficiency, we define the
\emph{equivalent diffusivity}\footnote{We refrain from calling~$\DiffEq$ an
``effective'' diffusivity because this already carries a definition in the
literature (\eg \cite[Young 1999]{GFD1999}).  There the effective
diffusivity is defined in terms of a large-scale gradient in the
concentration, whereas here we use the amplitude of the source, which makes
more sense in the present context.  The relationship between that 
traditional effective diffusivity~$\DiffEff$ and~$\DiffEq$ 
is~$\DiffEq=\DiffEff\,(\Theta/G\ell)^2$, where~$G$ is a linear gradient of
concentration (\cite[Schumacher, Sreenivasan \& Yeung 2003]{Schumacher2003}).
Other notions of effective diffusivity are also used in the context of
anomalous diffusion (\eg \cite[Isichenko 1992]{Isichenko1992}) and turbulence
(\eg \cite[Pope 2000]{Pope}).}
\begin{equation}
  \DiffEq \ldef \beta\,\frac{\Ssc\ell^2}{\Theta}
  \le \ccbi\,U\ell + \ccbii\,\Diff,
  \label{eq:DiffEqdef}
\end{equation}
The factor~$\beta$ is the norm of the solution of the purely diffusive
problem,
\begin{equation}
  \beta \ldef
  \Lnorm{\l(\Sr\,\Pe\,\pd_\Time + \lapl_\yv\r)^{-1}\Phi}{L^2(\Unit^\sdim)}\,,
  \label{eq:betadef}
\end{equation}
and the constants~$\ccbi$ and~$\ccbii$ are respectively~$\cci$ and~$\ccii$
multiplied by~$\beta$.  The extra factor of~$\beta$ ensures
that~$\DiffEq=\Diff$ for~$U=0$, which is the purely diffusive case.  This
corresponds to the choice \hbox{$\Psi=(\Sr\,\Pe\,\pd_\Time +
\lapl_\yv)^{-2}\Phi/\beta$}, for which~$\ccbii=1$.  Note
that~$(\Sr\,\Pe\,\pd_\Time + \lapl_\yv)^{-N}$ is defined in the Galerkin sense
on the Fourier expansion of~$\Phi$.

The equivalent diffusivity $\DiffEq$ compares the source amplitude 
($\Ssc$) to the steady-state fluctuations in the concentration field 
($\Theta$); as its name implies, it may be regarded as the molecular 
diffusivity needed to give a comparable amount of mixing in the absence of 
flow.  
A high P\'{e}clet number steady-state mixing device should operate with as 
high an equivalent diffusivity as possible compared to the molecular 
diffusivity.
Alternatively, we may interpret the ratio $\DiffEq^2/\Diff^2$ as the  
supression factor for the solution's variance.  
That is, if $\theta_0$ is the solution of the diffusion equation with the 
same source but no stirring and $\Theta_0^2$ is it's variance, then the 
definition (\ref{eq:DiffEqdef}) is simply $\DiffEq^2/\Diff^2 = 
\Theta_0^2/\Theta^2$.

In the regime of small~$U$, the variance is proportional to the amplitude of
the source, a response we expect when the stirring does not play an important
role.  A large equivalent diffusivity means that we are getting a well-mixed
distribution (small~$\Theta$) compared to the initial inhomogeneity in the
source ($\Ssc$); as explained in Section~\ref{sec:bounds}, for fixed~$\Diff$
and~$\ell$ this can only be achieved by increasing~$U$.

The equivalent diffusivity can also be bounded from below by
using~\eqref{eq:Ssclowerbound},
\begin{equation}
  \DiffEq \ge \Diff\,\beta\l(2\pi\,\ell/L\r)^2\,.
  \label{eq:DiffEqlowerbound}
\end{equation}
The worst lower bound for the mixing efficiency would be achieved by injecting
scalar variance at scale~$\ell$ while stirring to keep the dominant scale of
the concentration fluctuation field as~$L$.

\section{Bounds for the Sine Flow}
\label{sec:sineflow}

As an example application, we consider the well-studied two-dimensional
Zeldovich sine flow, or random wave flow (\cite[Pierrehumbert
1994]{Pierrehumbert1994}; \cite[Antonsen \etal\ 1996]{Antonsen1996}).  This
flow consists of
\begin{figure*}
\centering
\psfrag{y1}{$y_1$}
\psfrag{y2}{\raisebox{.75em}{$y_2$}}
\psfrag{u1}{$\uv^{(1)}$}
\psfrag{u2}{$\uv^{(2)}$}
\subfigure[]{
	\includegraphics[width=.45\textwidth]{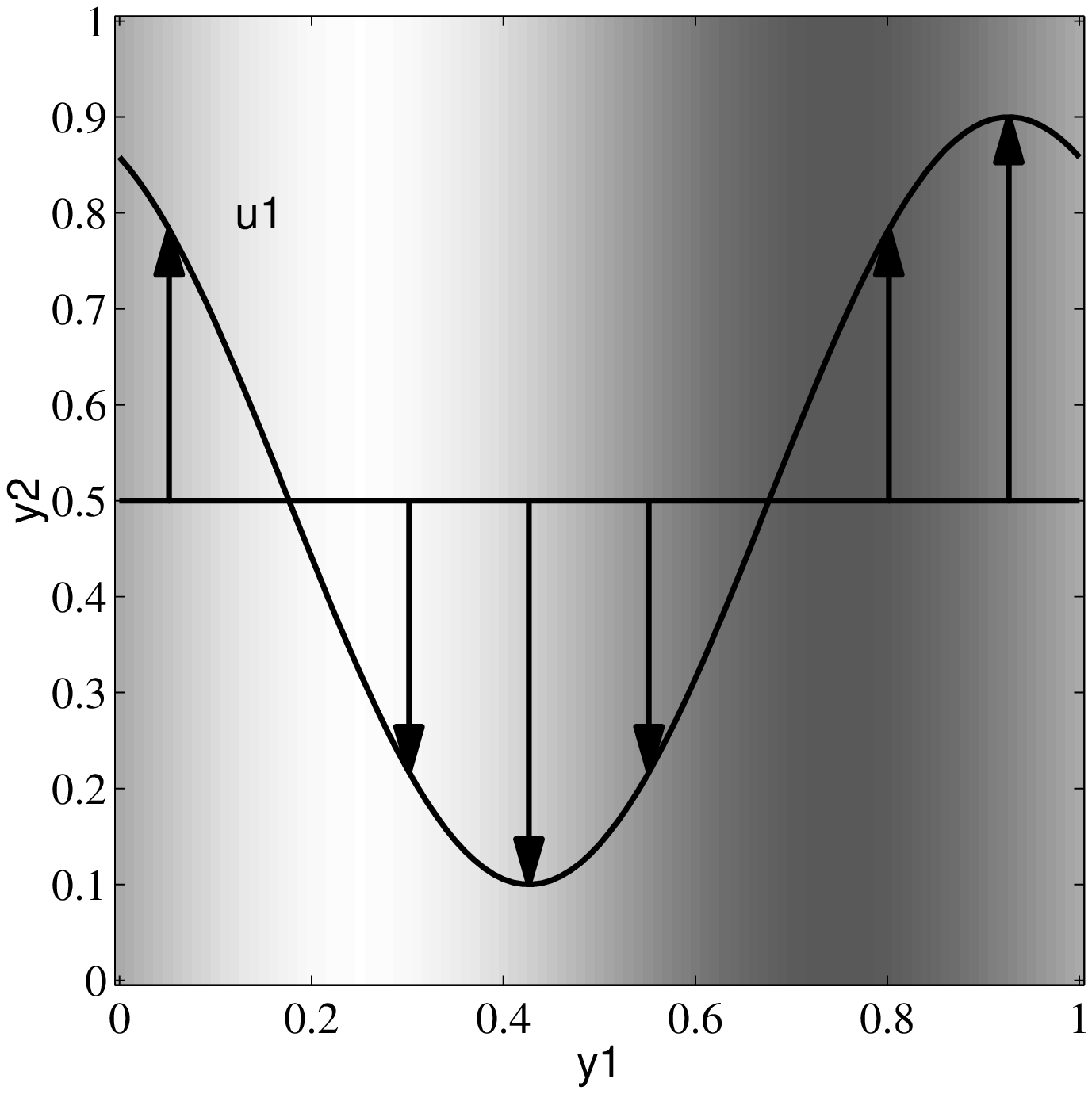}
	\label{fig:sineflow1}
}\goodgap
\subfigure[]{
	\includegraphics[width=.45\textwidth]{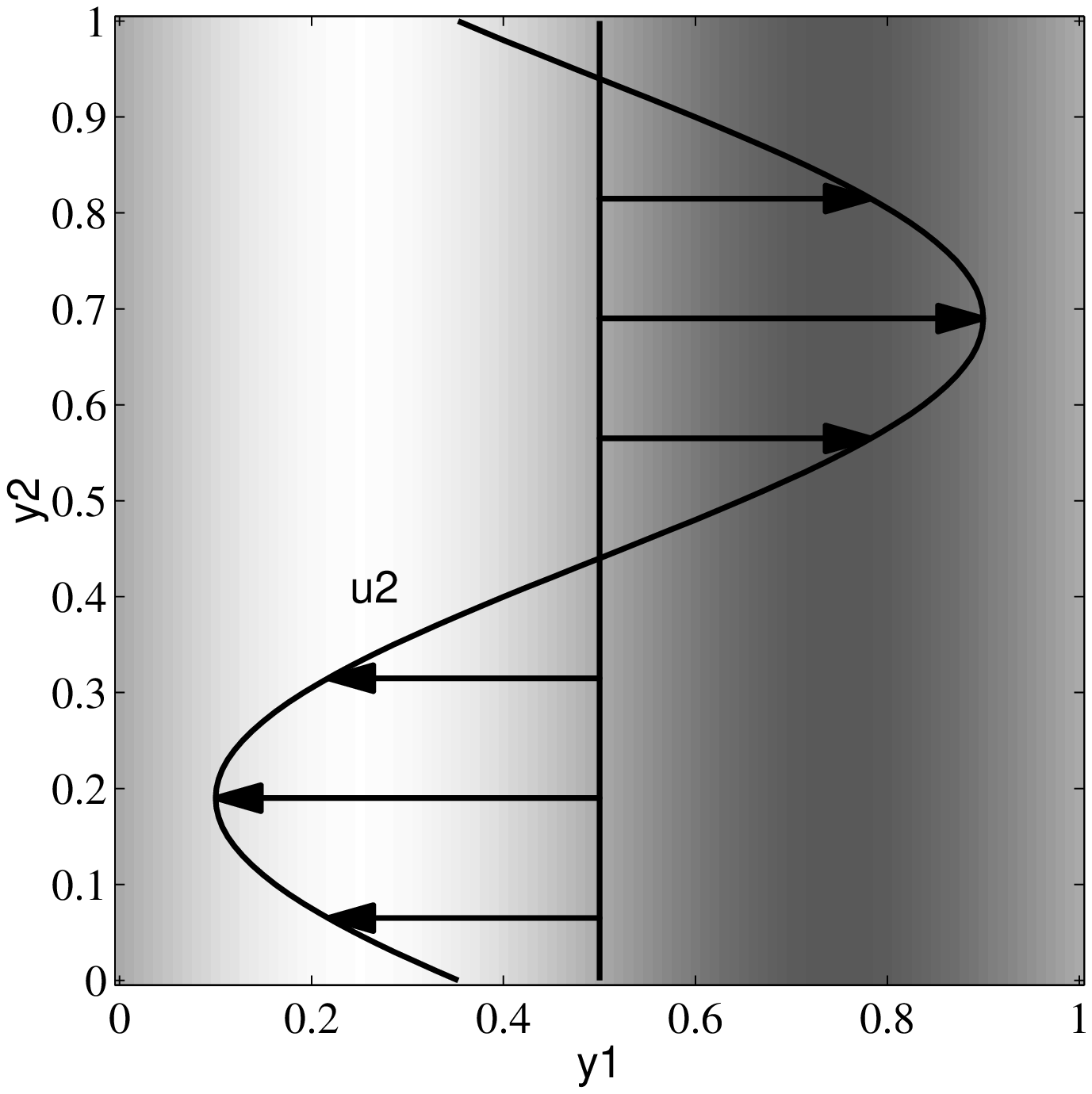}
	\label{fig:sineflow2}
}
\caption{The sine flow~\eqref{eq:sineflow} for (a) the first half and (b) the
  second half of each period, represented here with random phases~$\chi_1$
  and~$\chi_2$. The velocity field alternates direction, but the source
  distribution (as represented by the shaded background) is stationary.}
\label{fig:sineflow}
\end{figure*}
alternating horizontal and vertical sine shear flows, with phase
angles~$\chi_1$ and~\hbox{$\chi_2 \in [0,2\pi]$} randomly chosen at each time
period,~$\tau$ (see Figure~\ref{fig:sineflow}).  In the first half of the
period, the velocity field is
\begin{subequations}
\begin{equation}
  \uv^{(1)}(\xv,\time)
  = \sqrt{2}\, U \l(0\ ,\ \sin(2\pi x_1/L + \chi_1)\r);
    \label{eq:sineflow1}
\end{equation}
and in the second half-period it is
\begin{equation}
  \uv^{(2)}(\xv,\time)
  = \sqrt{2}\, U \l(\sin(2\pi x_2/L + \chi_2)\ ,\ 0\r).
    \label{eq:sineflow2}
\end{equation}
\label{eq:sineflow}%
\end{subequations}
\noindent
The flow is incompressible, and~$U$ is defined consistently
with~\eqref{eq:SUdef}, so that~$\Upsv$ is read off from~\eqref{eq:sineflow} by
dropping~$U$ and replacing~$\xv/L$ by~$\yv$.  As a source function, we choose
\hbox{$\ssc(\xv) = \sqrt{2}\,\Ssc\,\sin(2\pi x_1/L)$}, from
which~\hbox{$\Phi(\yv) = \sqrt{2}\,\sin(2\pi y_1)$}.  Here the source and
stirring scale length~$\ell$ is equal to the system size~$L$.  The
purely-diffusive solution with this source distribution
gives~$\beta=1/(2\pi)^2$ in Eq.~\eqref{eq:betadef}, and hence the lower
bound~\hbox{$\DiffEq \ge \Diff$} for the equivalent diffusivity.

The challenge now lies in choosing~$\Psi$ to optimise the bound as best we
can.  The simplest choice is to take~$\Psi=\Phi$, as this automatically
satisfies the normalisation~\eqref{eq:Psinorm}.  Inserting that form
into~\eqref{eq:c12def} (with~$\pd_\Time\Psi=0$), we find~\hbox{$\cci =
\sqrt{2}\,\pi$} and~\hbox{$\ccii = (2\pi)^2$}, for a bound on the equivalent
diffusivity
\begin{equation}
  \frac{\DiffEq}{\Diff} \le \frac{\Pe}{2\sqrt{2}\pi} + 1\,.
  \label{eq:sinebound}
\end{equation}
We can get a tighter bound by using~\eqref{eq:Sscbound}, which doesn't use the
Minkowski inequality, and exploiting the statistical isotropy and homogeneity
of the flow:
\begin{equation}
  \frac{\DiffEq}{\Diff} \le \sqrt{\frac{\Pe^2}{8\pi^2} + 1}\ .
  \label{eq:sineboundopt}
\end{equation}
The bound~\eqref{eq:sineboundopt} is actually optimal over
time-independent~$\Psi$ for our choice of stirring and source shape functions.
Because~$\Upsv$ is discontinuous in time (which is not an obstacle to the
bounding procedure), this particular velocity field does not yield a form
of~$\Omega$ that is easily optimized over time-dependent~$\Psi$.  So for this
example with a steady source and a time independent multiplier~$\Psi$, our
bound is uniform in the Strouhal number $\Sr$.
\begin{figure}
\psfrag{Deff/D}{\raisebox{.2cm}{$\DiffEq/\Diff$}}
\psfrag{Pe}{\raisebox{-.2cm}{$\Pe$}}
\begin{center}
\includegraphics[width=.7\columnwidth]{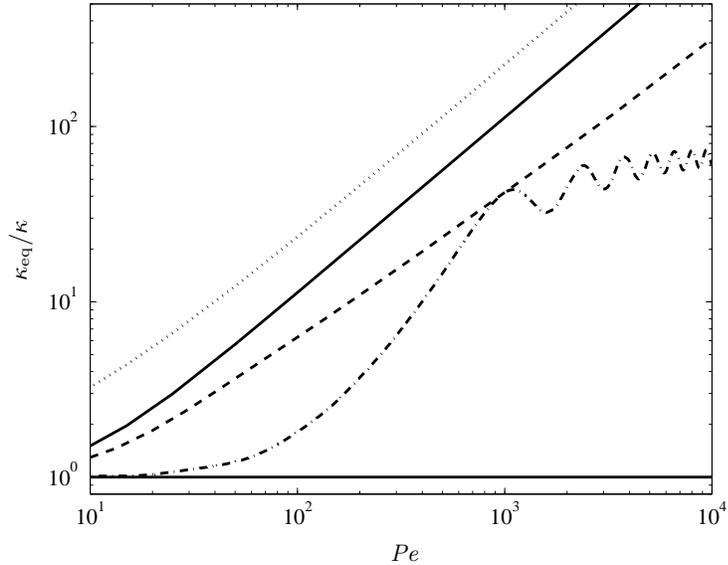}
\end{center}
\caption{Ratio of equivalent diffusivity~$\DiffEq$ to molecular
diffusivity~$\Diff$ for the sine flow~\eqref{eq:sineflow}.  The upper solid
line is the upper bound~\eqref{eq:sineboundopt} and the lower
bound~\eqref{eq:DiffEqlowerbound} is shown as a horizontal solid line.  The
upper limit~\eqref{eq:absbound2} indicated by the dotted line is valid for
{\it any} stirring flow with this source distribution.  The dashed line is the
result of direct numerical simulations with~$U$ and~$\tau$ fixed (\ie, with
constant~$\Sr=1$).  The dashed-dot curve plots simulation data with $\Diff$
and $\tau$ held constant while varying~$U$ (in this case $\Sr=\Pe^{-1}$).}
\label{fig:rwbound}
\end{figure}

Figure~\ref{fig:rwbound} shows the upper and lower bounds together with the
results of direct numerical simulations of the advection--diffusion
equation~\eqref{eq:AD} with this single-mode source and the sine
flow~\eqref{eq:sineflow}.  The upper and lower bounds~\eqref{eq:sineboundopt}
and~\eqref{eq:DiffEqlowerbound} are plotted as solid lines.

There are various ways of varying~$\Pe$ in this model.  For one, we can hold
$U$ and $\tau$ fixed (both at the value 1) and vary $\Diff$, in which case the
Strouhal number is fixed.  The dashed line in Figure~\ref{fig:rwbound} is the
result of the simulation with $\Sr=1$ (in all simulations reported here,
$\ell=L=1$).  We see that the bound qualitatively captures the behaviour of
the equivalent diffusivity in this case, although we do not have a wide enough
high $\Pe$ range to determine if the high-$\Pe$ asymptotic scaling agrees with
the bounds'.

Another simulation strategy is to hold $\Diff$ and $\tau$ fixed (in this case
at $10^{-3}$ and $1$, respectively) and vary $U$.  The data from this
simulation with $\Sr=\Pe^{-1}$ are plotted as the dash-dot line in
Figure~\ref{fig:rwbound}.  We observe that this method of stirring is less
effective at suppressing variance in the concentration at any value of~$\Pe
\ne 10^3$ where all the parameters coincide.  In particular, with this
stirring the enhancement of the equivalent diffusivity tends to saturate
rather than increase indefinitely at high~$\Pe$.  This can be understood as a
``resonance'' effect of the periodic boundary conditions: as $U$ increases at
fixed $\tau$ and the typical displacement $U\tau$ exceeds $\ell$, the velocity
field merely maps the concentration fluctuations onto periodic copies of
themselves rather than mixing it up within each periodic cell.  Because the
bound in~\eqref{eq:sineboundopt} is uniform in $\tau$, it should be compared
at a given value of~$\Pe$ against the largest possible equivalent diffusivity
achievable by any $\tau$, which can only improve the agreement with the bound.
These different simulation schemes illustrate the importance of the Strouhal
number for the mixing efficiency and, not unexpectedly, highlight the need for
further analysis to extract the Strouhal number dependence of the best bounds.

The dotted line in Figure~\ref{fig:rwbound} is a weaker upper bound obtained
from~\eqref{eq:absbound2} using~$\Psi=\Phi$; it sets an absolute limit on the
mixing efficiency achievable with \emph{any} stirring field shape at any
Strouhal number for this particular source distribution.

Finally, we note that there are flow fields at arbitrarily high values of
$\Pe$ and arbitrary $\Sr$ that saturate the {\it lower} bound $\DiffEq/\Diff =
1$ for this source shape.  Indeed, any flow field $\uv$ with no $x_2$
dependence (and arbitrary $x_1$ and $t$ dependence) simply moves the scalar
along iso-concentration lines to no effect whatsoever.  This simple example is
a particular case of a more general result concerning the existence of
``ineffective'' stirring fields (\cite[Young 2004]{WRYpc})---essentially
integrable fields without chaos.

\section{Conclusions}

It is encouraging that the equivalent diffusivities in
Figure~\ref{fig:rwbound} rise away from the diffusive lower bound as $\Pe$
increases, indicating that there is hope of more nearly saturating the upper
bound with more complex flows.  From the example of the sine flow, it is clear
that in general there is a nontrivial $\Sr$ dependence, even for steady
sources, that warrants further investigation.  Unlike the solution of the full
problem which requires a nonzero diffusivity to keep~$\Theta$ uniformly
bounded in time, the bounding procedure does not require any diffusivity.
That is, for large~$\Pe$ we may neglect~$\ccii$ from the bound altogether and
focus on~$\cci$ to try and minimize it with respect to~$\Psi$.  Of course, the
resulting optimal bound on $\DiffEq/U\ell$ may still depend in a complicated
way on~$\Pe$ and $\Sr$ for specific stirring and source distributions.

The high-$\Pe$ scaling of the bound obtained in this paper might be related to
an analogous one in combustion theory (\cite[Constantin \etal\
2000]{Constantin2000}).  There it was found that the bulk burning rate $V$ can
satisfy an ``optimal linear enhancement bound,'' $V \ge KU$, where $K$ is a
constant and $U$ is the magnitude of the advecting field.  The type of flow
required for linear enhancement, called ``percolating flows'' in
\cite{Constantin2000}, connects distant regions of unburned material.  Perhaps
these flows also produce linear asymptotic scaling with~$\Pe$ for the
equivalent diffusivity enhancement, but we have not yet investigated this.

Although we specified a body source in our problem with periodic conditions, a
source of concentration at impenetrable boundaries can be mimicked by a sharp
source concentrated near the walls (\cite[Balmforth \& Young
2003]{Balmforth2003}).  However, the type of wall boundary condition that can
be modelled in this manner is restricted to fixed scalar flux.

In closing we note that all of our analysis, as well as the general result
that $\DiffEq/\Diff \ldef \beta\Ssc\ell^2/\Diff\Theta \le \ccbi\Pe + \ccbii$,
depend on the source distribution being smooth enough to have a finite
variance $S^2$.  Point sources, for example where $\ssc \sim \delta(\xv)$, may
be of interest in applications but do not have finite variance.  In this
situation we may still define the mixing efficiency and an equivalent
diffusivity via $\DiffEq/\Diff \ldef \Theta_0/\Theta$ where $\Theta^2$ and
$\Theta_0^2$ are the scalar variances with and without the stirring; these
scalar variances are finite even for $\delta$-like sources in two and three
spatial dimensions.  However the anticipated behaviour suggested by the
consideration of smooth sources, i.e., that the equivalent diffusivity
enhancement $\DiffEq/\Diff$ and/or its upper bound could be $\sim \Pe$, may
not be realized with more singular sources.  The investigation of those models
is left for future work.

\begin{acknowledgments}

We thank P. Constantin, K.~R. Sreenivasan, and W.~R. Young for helpful
comments.  J.-L.T. and C.R.D. are grateful for the hospitality of the 2002
Summer Program in Geophysical Fluid Dynamics at the Woods Hole Oceanographic
Institution, where this work was initiated.  C.R.D. was supported in part by
NSF Awards PHY9900635 and PHY0244859.

\end{acknowledgments}


\begin{thebibliography}{}

\bibitem[{Antonsen, Jr.} \etal\ (1996)]{Antonsen1996}
{\sc {Antonsen, Jr.}, T.~M., Fan, Z., Ott, E. \& Garcia-Lopez, E.}, 1996.
\newblock The role of chaotic orbits in the determination of power spectra.
\newblock {\em Phys. Fluids\/} {\bf {\bf 8}(11)}, 3094--3104.

\bibitem[Aref (1984)]{Aref1984}
{\sc Aref, H.}, 1984.
\newblock Stirring by chaotic advection.
\newblock {\em J. Fluid Mech.} {\bf 143}, 1--21.

\bibitem[Balmforth \& Young (2003)]{Balmforth2003}
{\sc Balmforth, N.~J. \& Young, W.~R.}, 2003.
\newblock Diffusion-limited scalar cascades.
\newblock {\em J. Fluid Mech.} {\bf 482}, 91--100.

\bibitem[Batchelor (1959)]{Batchelor1959}
{\sc Batchelor, G.~K.}, 1959.
\newblock Small-scale variation of convected quantities like
temperature in turbulent fluid.
\newblock {\em J. Fluid Mech.} {\bf 5}, 113.

\bibitem[Childress \& Gilbert (1995)]{STF}
{\sc Childress, S. \& Gilbert, A.~D.}, 1995.
\newblock {\em Stretch, Twist, Fold: The Fast Dynamo\/}.
\newblock Springer-Verlag, Berlin.

\bibitem[Constantin \etal\ (2000)]{Constantin2000}
{\sc Constantin, P., Kiselev, A., Oberman, A. \& Ryzhik, L.}, 2000.
\newblock Bulk burning rate in passive--reactive diffusion.
\newblock {\em Arch. Rational Mech. Anal.} {\bf 154}, 53--91.

\bibitem[Danckwerts (1952)]{Danckwerts1952}
{\sc Danckwerts, P.~V.}, 1952.
\newblock The definition and measurement of some characteristics of mixtures.
\newblock {\em Appl. Sci. Res. A} {\bf 3}, 279--296.

\bibitem[Doering \& Foias (2002)]{Doering2002b}
{\sc Doering, C.~R. \& Foias, C.}, 2002.
\newblock Energy dissipation in body-forced turbulence.
\newblock {\em J. Fluid Mech.} {\bf 467}, 289-306.

\bibitem[Doering, Eckhart \& Schumacher (2003)]{Doering2003}
{\sc Doering, C.~R., Eckhart, B. \& Schumacher, J.}, 2003.
\newblock Energy dissipation in body-forced plane shear flow.
\newblock {\em J. Fluid Mech.} {\bf 494}, 275--284.

\bibitem[Edwards, Sherman \& Breidenthal (1985)]{Edwards1985}
{\sc Edwards, A.~C., Sherman, W.~D. \& Breidenthal, R.~E.}, 1985.
\newblock Turbulent mixing in tubes with transverse injection.
\newblock {\em AIChE J.} {\bf 31}, 516.

\bibitem[Isichenko (1992)]{Isichenko1992}
{\sc Isichenko, M.~B.}, 1992.
\newblock Percolation, statistical topography, and transport in
random-media.
\newblock {\em Rev. Mod. Phys.} {\bf 64}, 961--1043.

\bibitem[Ottino (1989)]{Ottino}
{\sc Ottino, J.~M.}, 1989.
\newblock {\em The Kinematics of Mixing: Stretching, Chaos, and
 Transport\/}.
\newblock Cambridge University Press, Cambridge, U.K.

\bibitem[Pierrhumbert (1994)]{Pierrehumbert1994}
{\sc Pierrehumbert, R.~T.}, 1994.
\newblock Tracer microstructure in the large-eddy dominated regime.
\newblock {\em Chaos Soitons Fractals} {\bf 4}, 1091--1110.

\bibitem[Pope (2000)]{Pope}
{\sc Pope, S.~B.}, 2000. \newblock {\em Turbulent Flows\/}.
\newblock Cambridge University Press, Cambridge, U.K.

\bibitem[Rehab, Antonia, Djenidi \& Mi (2000)]{Rehab2000}
{\sc Rehab, H., Antonia, R.~A., Djenidi, L. \& Mi, J.}, 2000.
\newblock Characteristics of fluorescein dye and temperature fluctuations in a
turbulent near-wake.
\newblock {\em Exp. Fluids} {\bf 28}, 462--470.

\bibitem[Schumacher, Sreenivasan \& Yeung (2003)]{Schumacher2003}
{\sc Schumacher, J., Sreenivasan, K.~R. \& Yeung, P.~K.}, 2003.
\newblock Schmidt number dependence of derivative moments for quasi-static
straining motions.
\newblock {\em J. Fluid Mech.} {\bf 479}, 221--230.

\bibitem[Young (1999)]{GFD1999}
{\sc Young, W.~R.}, 1999.
\newblock {\em Stirring and Mixing: Proceedings of the 1999 Summer Program in
  Geophysical Fluid Dynamics\/}, edited by J.-L.~Thiffeault and C.~Pasquero.
\newblock Woods Hole Oceanographic Institution, Woods Hole, MA, USA.
\newblock
  \texttt{http://gfd.whoi.edu/proceedings/1999/PDFvol1999.html}

\bibitem[Young (2004)]{WRYpc}
{\sc Young, W.~R.}, 2004.
\newblock Private communication.


\end{thebibliography}

\end{document}